# Exchange Frustration and Topological Magnetism in Electrostatically Doped SrRuO$_3$

Naafis Ahnaf Shahed, Himanshu Mavani, Zhonglin He, Kai Huang, Mohamed Elekhtiar, and Evgeny Y. Tsymbal

*Department of Physics and Astronomy & Nebraska Center for Materials and Nanoscience,*
*University of Nebraska, Lincoln, Nebraska 68588-0299, USA*

Magnetism in transition-metal systems emerges from exchange interactions that depend sensitively on carrier density. Yet leveraging this sensitivity to deliberately engineer exchange frustration and associated topological spin textures remains largely unexplored. Here, combining first-principles calculations with atomistic Monte Carlo simulations, we demonstrate that ferroelectric polarization enables electrostatic control of exchange frustration in the itinerant ferromagnet SrRuO$_3$. We show that electrostatic hole doping renormalizes competing exchange interactions, driving SrRuO$_3$ away from its bulk ferromagnetic ground state toward frustrated regimes, whereas electron doping largely preserves ferromagnetism. At BaTiO$_3$/SrRuO$_3$ interfaces, polarization-induced charge depletion modulates layer-dependent exchange couplings, enhancing competition among $J_1$, $J_2$, and $J_3$. The resulting exchange frustration stabilizes a sequence of magnetic phases as a function of thickness and applied magnetic field, including stripe and spiral states, topological meron and bimeron textures, and diverse skyrmionic objects. A minimal spin model identifies exchange frustration as the primary control parameter governing these crossovers, with magnetic anisotropy, Dzyaloshinskii-Moriya interaction, and external field selecting the emergent topology. Our results establish electrostatic doping as a route to engineer frustrated and topological magnetism in itinerant oxide metals.

Magnetism in transition-metal systems is governed by exchange interactions whose magnitude and sign depend on band filling. As the occupation of partially filled $d$ bands is varied, the dominant exchange coupling can reverse sign, leading to transitions between ferromagnetic (FM) and antiferromagnetic (AFM) alignments. In localized and mixed-valence oxides, this behavior arises from the competition between AFM super-exchange, described by Anderson and the Goodenough-Kanamori rules [1,2], and FM kinetic double-exchange introduced by Zener and formalized by Anderson and Hasegawa [3,4]. In itinerant-electron systems, exchange interactions are controlled by the spin-dependent band structure and the underlying magnetic susceptibility. Changes in band filling shift the Fermi level and can drive sign reversal of effective exchange parameters within Stoner and spin-fluctuation descriptions of itinerant magnetism [5,6].

A classic realization of band filling-controlled magnetism occurs in doped perovskite manganites. In La$_{1-x}$Sr$_x$MnO$_3$ and related $R_{1-x}A_x$MnO$_3$ compounds, chemical substitution tunes the Mn valence and $e_g$ occupancy, driving a rich phase diagram that includes FM metallic phases, multiple AFM orders, and orbitally ordered insulating states [7–9]. The evolution of magnetic order with doping reflects the delicate balance between competing exchange interactions, bandwidth control, and lattice distortions.

Beyond chemical substitution, carrier density can also be tuned electrostatically, avoiding chemical disorder and structural changes. In ferroelectric/metal heterostructures, polarization of the ferroelectric layer induces screening charge in the adjacent metallic magnet, effectively acting as a reversible electrostatic doping mechanism. Polarization reversal modifies the occupation of interfacial $d$ states and can thereby alter magnetic anisotropy and exchange interactions [10–14].

While chemical and electrostatic doping have been widely used to control magnetism, their deliberate use to engineer magnetic frustration remains largely unexplored. Due to nearest-neighbor (NN) exchange interactions ($J_1$, $J_2$, $J_3$) exhibiting different dependencies on band filling, carrier doping can selectively renormalize competing exchange pathways and thereby modify the magnetic ground state. In regimes where these interactions strongly compete, magnetic frustration may stabilize noncollinear spin textures and emergent topological phases.

In this context, SrRuO$_3$ (SRO) provides an ideal platform for exploration. As a prototypical $4d$ itinerant FM perovskite oxide, SRO is widely used as a conductive electrode and exhibits excellent epitaxial compatibility with other perovskites [15–17]. Its transport and magnetic properties are highly sensitive to carrier density, strain, film thickness, and interfacial symmetry breaking, all of which modify its electronic structure [15,18–21]. Even modest perturbations can substantially alter SRO magnetism, including suppression of FM order in ultrathin films [19,22]. Interface-induced electronic reconstruction together with spin-orbit coupling and distortions of the RuO$_6$ octahedra can enhance magnetic frustration [23,24]. As a result, SRO-based heterostructures have exhibited signatures consistent with chiral magnetic interactions and topological Hall-like responses [25,26], including in ferroelectric/SRO systems where electrostatic doping effects play a decisive role [27–29].

Although these phenomena have been experimentally observed, a quantitative microscopic understanding of how electrostatic carrier modulation reshapes competing exchange interactions to stabilize noncollinear, chiral, or other topological magnetic states in itinerant oxides remain incomplete. Previous work has largely emphasized the influence of ferroelectric polarization on the Dzyaloshinskii-Moriya interaction (DMI) [27,30–33], whereas its impact on magnetic frustration has remained largely unexplored.

Motivated by this challenge, in this Letter, we demonstrate that electrostatic doping provides a direct route to access frustrated and topological magnetism at the SrRuO$_3$/BaTiO$_3$ (SRO/BTO) interface. Our calculations based on density functional theory (DFT) show that hole doping strongly affects the exchange interactions in SRO, inducing magnetic frustration and favoring noncollinear magnetic order, whereas electron doping preserves the FM ground state. Mapping the extracted exchange parameters onto atomistic spin models reveals noncollinear spin textures, including topological meron and bimeron states that evolve under applied magnetic fields. Thickness-dependent calculations show that reduced dimensionality enhances magnetic frustration, stabilizing stripe domains and spiral phases, whereas increasing thickness strengthens perpendicular anisotropy and favors field-stabilized skyrmionic textures.

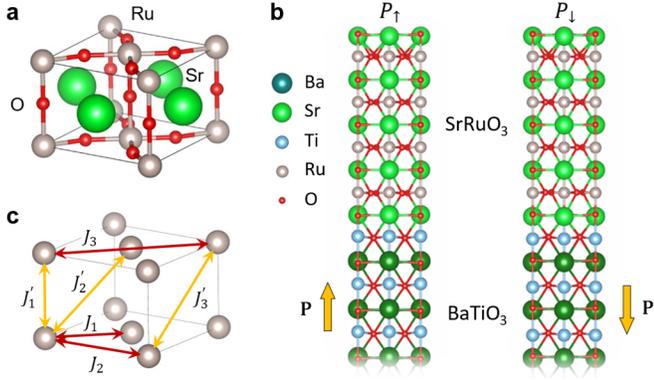

FIG. 1. Atomic structure and exchange parameters. (a) Atomic structure of the SRO unit cell in the tetragonal $P4/mbm$ phase. (b) Schematic of the SRO/BTO heterostructure showing two interfaces with opposite orientations of the BTO ferroelectric polarization relative to SRO, $P_\downarrow$ and $P_\uparrow$. (c) Exchange interactions between Ru moments, distinguishing in-plane nearest neighbors $J_m (m = 1,2,3)$ (blue arrows) and out-of-plane interactions $J'_m$ $(m = 1,2,3)$ (green arrows).

To analyze the effect of electrostatic doping on SRO, we perform DFT calculations using a slab supercell of vacuum/SRO/BTO/SRO/vacuum. SRO is assumed to adopt the tetragonal $P4/mbm$ phase [Figs. 1(a)], which is typical of epitaxial SRO films grown on $SrTiO_3$ (e.g., [34]). The in-plane lattice constant is fixed to that of $SrTiO_3$ to mimic epitaxial strain conditions. The supercell comprises 5.5-unit-cell (u.c.) BTO, $m$-u.c. SRO layers ($m = 2.5$–$5.5$), and a 20 Å vacuum region and has $TiO_2$-SrO termination at the SRO/BTO interface (see Supplementary Material [35] for computational details and structural model). Two symmetry-inequivalent interfaces arise due to opposite orientations of the BTO ferroelectric polarization relative to SRO, $P_\downarrow$ and $P_\uparrow$, as illustrated in Figs. 1(b) and S2.

We find that structural relaxation preserves ferroelectric distortions in BTO, with sizable Ti off-centering, $\Delta_{Ti} \sim 0.18$ Å, and a pronounced tetragonality, induced by ferroelectricity and compressive strain. These polar displacements penetrate the metallic SRO layer and decay away from the interface, thereby breaking inversion symmetry in SRO. In addition, the polarization of strained BTO ($P \approx 55$ μC/cm²) induces screening charge in the SRO layer, estimated to be $\sim 0.5$ $e$ per lateral unit cell, with negative (positive) sign for $P_\uparrow$ ($P_\downarrow$), corresponding to electrostatic electron (hole) doping. This is consistent with previous experimental findings, where strained BTO has exhibited polarization up to 70 μC/cm² [36].

This structural asymmetry and electrostatic doping alter the interfacial magnetic interactions, leading to variations in the Heisenberg exchange and magnetic anisotropy, as well as the emergence of DMI. We find that these effects differ markedly for the $P_\uparrow$ and $P_\downarrow$ interfaces. For the $P_\uparrow$ interface, corresponding to electron doping of SRO, the NN exchange interactions remain dominantly FM, stabilizing a FM ground state in the adjacent SRO layer, independent of the SRO thicknesses considered (Tables S3, S5, and S7 and Fig. S5 [35]). In contrast, for the $P_\downarrow$ interface, the exchange interactions compete with comparable magnitude. Therefore, the remainder of this work focus on the $P_\downarrow$ interface, where electrostatic hole doping of SRO stabilizes nontrivial magnetism.

We first consider SRO/BTO heterostructure with 2.5 u.c. of SRO. As evident from Table S2, the exchange interactions [Fig. 1(c)] become strongly layer dependent. While the surface Ru layer remains FM ($J_1 \approx 10$ meV), the interfacial layer exhibits a substantially reduced NN exchange ($J_1 \approx 3.3$ meV) together with sizable competing interactions ($J_2 \approx 1$ meV and $J_3 \approx -2.4$ meV). This combination leads to a large frustration parameter $|J_3/J_1|$, indicating proximity to a spiral instability [37]. In addition, a sizable DMI emerges, with a dominant out-of-plane component $D_z \sim 1$ meV that exceeds the local interfacial anisotropy energy and favors easy-plane magnetization.

Atomistic Monte Carlo simulations using these DFT parameters [35] demonstrate that the magnetic ground state is governed primarily by the frustrated exchange. As shown in Fig. 2(e), the enhanced frustration stabilizes stripe-like magnetic domains in both Ru layers. These domains represent helical spin spirals propagating along the [100] direction with a spiral period of about $\lambda \approx 18$ Å and magnetization rotating out of the film plane as it varies spatially.

The emergence of helical spin spirals in hole-doped SRO can be qualitatively understood using a minimal tight-binding model [35]. Because doping primarily affects the in-plane exchange interactions, while the out-of-plane exchange remains strong and maintains a nearly collinear alignment, we restrict the model to a 2D description of the $RuO_2$ planes. The model includes three $t_{2g}$-like orbitals ($d_{xy}$, $d_{xz}$, $d_{yz}$) on a square lattice with NN hopping $t = t_{xy}$ and $t' = t_{xz} = t_{yz}$, neglects inter-orbital hybridization, and assumes degenerate on-site energies [Fig. 3(a)]. Magnetism is introduced through an orbital-independent exchange splitting parameter $\Delta$.

Exchange interactions $J_m$ ($m = 1,2,3$) are extracted from band energies of collinear magnetic configurations [35] and plotted in Fig. 3(b), as a function of band filling $n$, for $\Delta = t$ and $t' = 0.5t$. As seen from Fig. 3(b), the NN exchange $J_1$ exhibits the strongest dependence on band filling, changing sign as $n$ varies and driving transitions between FM, spiral, and AFM regimes. While the longer-range exchanges $J_2$ and $J_3$ remain smaller, they provide competing interactions that stabilize spiral states in the regions where $J_1$ is suppressed. The dashed line indicates the band filling $n$ approximately corresponding to bulk SRO, which lies near the boundary between the spiral and FM regimes, suggesting that moderate hole doping can drive the system into a frustrated spiral phase. Despite its simplicity, the model captures the essential mechanism by which charge modulation drives SRO from a strongly FM state toward a regime of competing magnetic interactions.

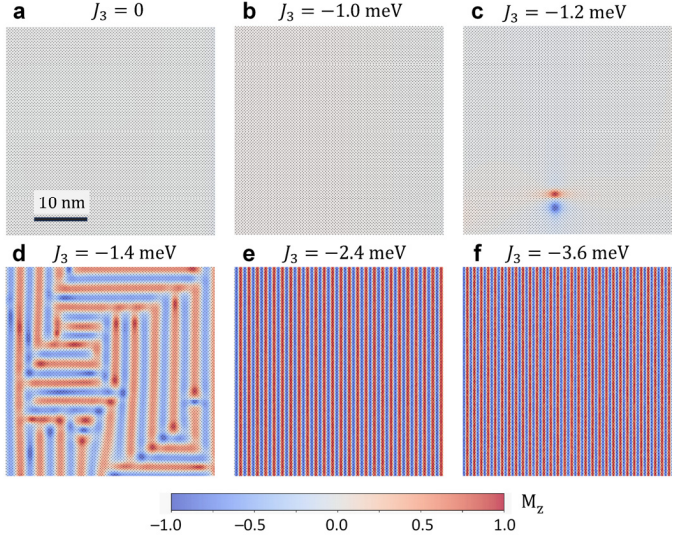

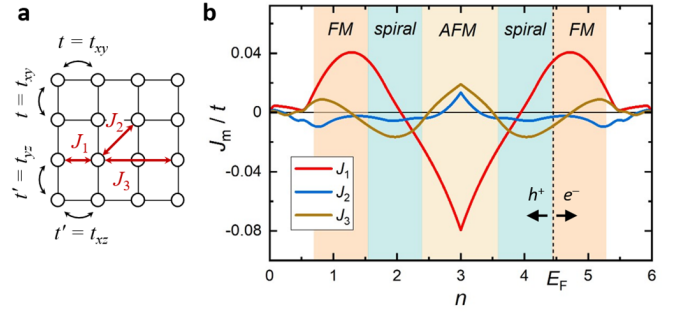

FIG. 2. Spin textures of 2.5-u.c. SRO/BTO heterostructure for different values of $J_3$. $J_3 = -2.4$ meV (e) corresponds to the value obtained from DFT. Other parameters are fixed obtained to those obtained for a 2.5-u.c. SRO/BTO $P_\downarrow$ interface layer: $J_1 = 3.3$ meV, $J_2 = 1.0$ meV, $D_x = D_y = 0.2$ meV, $D_z = 1.1$ meV, and $K = 0.04$ meV.

FIG. 3. Tight-binding (TB) model and its results. (a) Schematic of the TB model showing the hopping parameters of the $t_{2g}$ orbitals ($t$ and $t'$) and the NN exchange interactions $J_m$ ($m = 1,2,3$). (b) Exchange parameters $J_m$ as a function of band filling $n$, calculated with $\Delta = t$ and $t' = 0.5t$. Shaded regions indicate regimes where FM, AFM, or spiral magnetic order is stabilized. The vertical dashed line marks the approximate position of the Fermi level in bulk SRO, indicating that moderate electron ($e^-$) or hole ($h^+$) doping can drive a crossover between FM and spiral states.

Furthermore, the 2D model reproduces the key features of the spiral states obtained from DFT. Minimization of the classical spin energy yields a diagonal spiral with wave vector $\mathbf{q} = (q, q)$ satisfying $\cos q = -\frac{J_1}{2J_2 + 4J_3}$, with spiral period $\lambda = \frac{2\pi}{q}$ [35]. In this state the magnetization forms a helical spiral, rotating out of the layer plane as it propagates along the [100] direction. Using exchange parameters for the interface layer (layer 1) from Table S2, we estimate a spiral period of $\lambda \approx 22$ Å. This value is in qualitative agreement with the value obtained from DFT, indicating that the exchange frustration at the interface is mainly responsible for the formation of helical spin spirals (the layer-average values predict a much larger period of $\lambda \approx 31$ Å). Some discrepancy may reflect the influence of DMI neglected in the minimal model.

When $J_1 > 0$ (FM), $J_2$ is small, and $J_3 < 0$ (AFM), as in our case, the spiral period is primarily governed by $|J_3/J_1|$, reflecting the competition between FM NN exchange and AFM longer-range frustration. To investigate how the magnetic texture evolves with increasing frustration, we perform Monte Carlo simulations of a 2D system in which all magnetic parameters are fixed to those obtained for a 2.5-u.c. SRO/BTO $P_\downarrow$ interface and only $J_3$ is varied (Fig. 2).

As expected, for small $|J_3|$, the system remains essentially ferromagnetic [Figs. 2(a,b)]. With increasing frustration, the system approaches the Lifshitz point defined by $\left|\frac{J_1}{2J_2+4J_3}\right| = 1$, where the uniform FM state becomes unstable to finite-$q$ modulations [35]. In this intermediate regime, localized noncollinear textures emerge, including a bimeron-like state [Fig. 2(c)]. Upon further increasing $|J_3|$, the system evolves into extended spiral phases, first forming labyrinthine patterns [Fig. 2(d)] and then developing well-defined stripe-like spin spirals with progressively shorter wavelengths [Figs. 2(e,f)]. Thus, magnetic frustration acts as the control parameter governing the crossover from stripe magnetism to topological textures.

With increasing SRO thickness, the magnetic frustration, largely originating from the interface, is reduced, placing the system in the regime where topological quasiparticles can arise. As evident from Table S4, for a 4.5-u.c. SRO/BTO $P_\downarrow$ interface, the magnetic interactions exhibit a strong layer dependence. At the interfacial Ru layer, the in-plane first-NN exchange is strongly suppressed and becomes AFM ($J_1 = -2.8$ meV), while the longer-range couplings remain FM ($J_2 \approx 1.1$ meV and $J_3 \approx 0.6$ meV). In contrast, deeper Ru layers recover strong first-NN FM exchange ($J_1 \sim 7$–$10$ meV), whereas the third-NN exchange remains sizable and AFM ($J_3 \sim -1$–$4$ meV). A substantial DMI, $D_z \sim 1$ meV, persists across the entire layer thickness. In the presence of DMI, such frustrated exchange interactions are expected to stabilize topological magnetic textures (e.g., [38]).

This expectation is confirmed by our atomistic Monte Carlo simulations. The resulting magnetic configuration at zero field contains coexisting isolated merons and bimerons distributed randomly across the magnetic layers of the film [Fig. 4(a)].

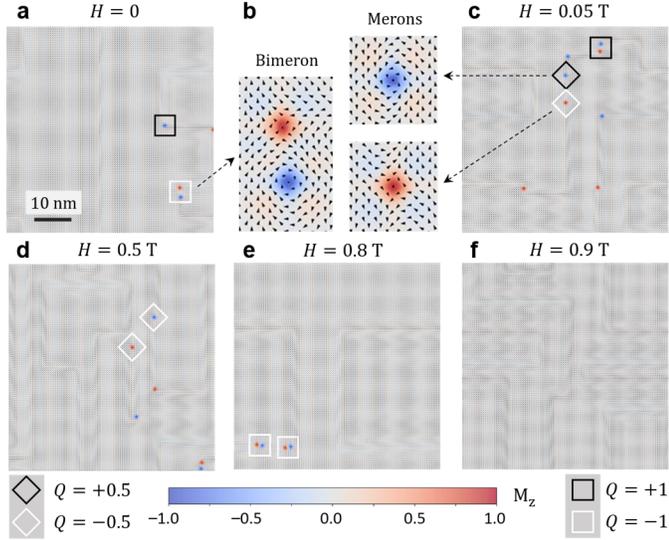
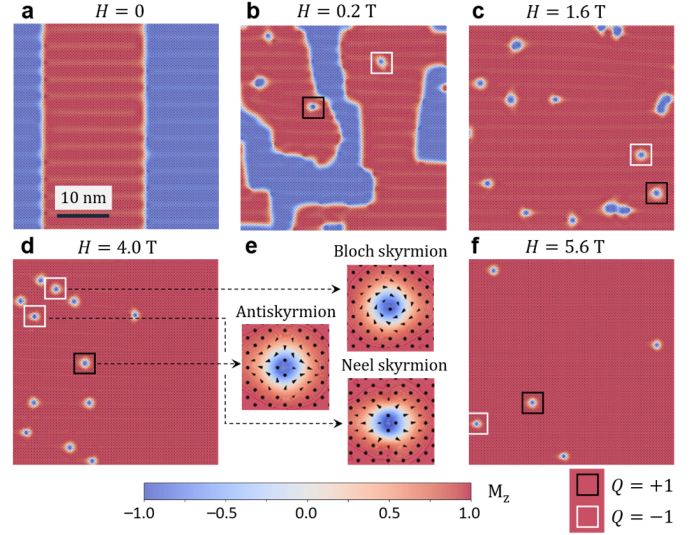

FIG. 4. Spin textures for 4.5-u.c. SRO/BTO heterostructure at different magnetic fields. (a) Zero-field magnetic states for the surface Ru layer. (c–f) Evolution of the spin configuration with increasing in-plane magnetic field $H$ applied along the [100] direction under field-cooling protocol. (b) Zoomed in texture of merons and a bimeron. Topological charge $Q$ of selected magnetic quasiparticles is indicated.

FIG. 5. Spin textures for 5.5-u.c. SRO/BTO heterostructure at different magnetic fields. (a) Nearly collinear multidomain ground state with chiral domain walls at zero field. (b–d,f) Evolution under an out-of-plane magnetic field $H$ showing nucleation of isolated skyrmions. (e) Zoomed in texture of Bloch and Néel skyrmions, and an antiskyrmion. Topological charge $Q$ of selected magnetic quasiparticles is indicated.

Merons represent in-plane topological textures carrying a fractional topological charge $Q = \pm 1/2$ [Fig. 4(b), right]. A bimeron can be viewed as a bound state of two merons carrying the same topological charge, forming a composite texture with integer topological charge $Q = \pm 1$ [Fig. 4(b), left], typically stabilized by frustrated exchange interactions [39,40]. In our system, meron textures emerge from the superposition of two orthogonal finite-$q$ modulations, while bimerons correspond to pairs of like-charged merons bound into a single object within an easy-plane magnetic background [41].

We note that the strong interlayer exchange interaction (> 10 meV) enforces nearly collinear alignment of spins across adjacent Ru monolayers. As a result, any noncollinearity developing within an SRO plane propagates coherently throughout the SRO layer. In the 4.5-u.c.-thick SRO film, where the two central monolayers are AFM coupled, the top and bottom pairs of Ru layers carry antiparallel magnetic moments, producing identical spin textures with opposite chirality in the two halves of the film (Fig. S6).

These textures evolve gradually under a field-cooling protocol with an in-plane magnetic field $H$ applied along the [100] direction [Figs. 4(c-f)]. At low fields, the system exhibits a multidomain noncollinear state, where meron- and bimeron-like textures are primarily localized at chiral domain boundaries, similar to the zero-field case, but with an increased density of topological objects up to $H \sim 0.3$ T. These defects emerge within a labyrinthine domain pattern, reflecting the competition between frustrated exchange interactions and field-induced alignment.

With increasing field, the domain walls become progressively less entangled, accompanied by a reduction in the density of topological textures. Above $H \sim 0.8$ T, meron- and bimeron-like textures disappear, and the system approaches a uniformly magnetized state. The emergence of merons and bimerons (and potentially higher-order topological textures) is consistent with theoretical predictions for frustrated easy-plane magnets near a Lifshitz point, where competing exchange interactions stabilize fractional topological defects and their bound states [39].

Next, we examine a thicker 5.5-u.c. SRO film, where increased dimensionality brings the system closer to bulk-like SRO, exhibiting out-of-plane magnetic anisotropy consistent with experiment [27]. In this case, the zero-field ground state remains nearly collinear and forms magnetic domains, with frustration and DMI playing a secondary role [Fig. 5(a)]. Under an out-of-plane magnetic field, however, the competition between Zeeman energy, exchange interactions, and DMI stabilizes noncollinear skyrmionics spin textures [Figs. 5(b–d,f)]. In this regime, we observe the formation of diverse magnetic quasiparticles, including Bloch- and Néel-type skyrmions as well as antiskyrmions, reflecting the rich interplay of competing interactions [Fig. 5(e)]. This behavior reflects a dimensional crossover from frustration-dominated easy-plane to easy-axis topology, consistent with general expectations [42]. We also observe larger skyrmionic objects with higher-order topological charge during the intermediate regime where chiral domain walls transform into isolated skyrmions [Fig. 5(c)]. All skyrmionic textures disappear at fields above 6.5 T.

Finally, we emphasize that the magnetic behavior observed at ferroelectric interfaces reflects an intrinsic property of *bulk* SrRuO$_3$: the strong sensitivity of its exchange interactions to carrier density. Our calculations show that, while electron doping preserves the FM ground state, hole doping strongly suppresses $J_1$ and even make it AFM (Table S1 [35]), consistent with our simple tight-binding model [Fig. 3(b)]. The resulting competition between $J_1$ and $J_3$ promotes magnetic frustration, providing a microscopic basis for the polarization-controlled magnetism predicted at ferroelectric interfaces.

In conclusion, we have demonstrated that electrostatic hole doping induced at BaTiO$_3$/SrRuO$_3$ interfaces gives rise to an unusually rich magnetic landscape in itinerant SrRuO$_3$, where competing exchange interactions drive the system away from its bulk FM state toward regimes of strong magnetic frustration. As a result, a variety of noncollinear magnetic states can emerge depending on thickness and applied magnetic field, including spiral and stripe phases as well as topological textures such as merons, bimerons, and diverse skyrmionic configurations. Our calculations capture the field-stabilized skyrmions observed experimentally [27]; however, the behavior may be significantly richer, warranting further investigation.

The tunable hierarchy of magnetic states opens several experimental avenues for exploration. For example, spin-resolved imaging techniques, such as Lorentz TEM, spin-polarized STM, or NV-diamond magnetometry, could directly probe the predicted stripe and topological textures, while transport measurements sensitive to emergent electrodynamics, such as the topological Hall effect, could provide signatures of bimeron or skyrmion quasiparticles. More broadly, electrostatic control of exchange frustration offers a promising route toward electrically tunable topological magnetism in itinerant oxide heterostructures.

This work was supported by the National Science Foundation under Award No. DMR-2522667 (atomic structure optimization, electronic structure calculations, tight-binding modeling), the U.S. Department of Energy, Office of Science, Basic Energy Sciences under Award No. DE-SC0026103 (evaluation of magnetic parameters and Monte Carlo simulations), and UNL Grand Challenges catalyst award "Quantum Approaches Addressing Global Threats" (computing topological charge). Computations were performed at the University of Nebraska Holland Computing Center.